**Highly stable photoluminescence in vacuum-processed halide perovskite core-shell 1D nanostructures**

*Javier Castillo-Seoane, Lidia Contreras-Bernal,\* T. Cristina Rojas, Juan P. Espinós, Andrés-Felipe Castro-Méndez, Juan-P. Correa-Baena, Angel Barranco, Juan R. Sanchez-Valencia,\* and Ana Borras*

Javier Castillo-Seoane, Lidia Contreras-Bernal, Juan Pedro Espinós, Ángel Barranco, Juan Ramon Sanchez-Valencia, Ana Borrás.
Nanotechnology on Surfaces and Plasma Lab, Institute of Materials Science of Seville (US-CSIC), Americo Vespucio 49, 41092 Seville, Spain
Teresa Cristina Rojas
Institute of Materials Science of Seville (US-CSIC), Americo Vespucio 49, 41092 Seville, Spain
E-mail: jrsanchez@icmse.csic.es; lidia.contreras@icmse.csic.es

Andrés-Felipe Castro-Méndez and Juan-Pablo Correa-Baena.
School of Materials Science and Engineering, Georgia Institute of Technology, Atlanta, GA, USA

Hybrid organometal halide perovskites (HP) present exceptional optoelectronic properties, but their poor long-term stability is a major bottleneck for their commercialization. Herein, we present a solvent-free approach to growing single-crystal organic nanowires (ONW), nanoporous metal oxide scaffolds, and HP to form a core@multishell architecture. The synthetic procedure is carried out under mild vacuum conditions employing thermal evaporation for the metal-free phthalocyanine ($H_2Pc$) nanowires, which will be the core, plasma-enhanced chemical vapor deposition (PECVD) for the $TiO_2$ shell, and co-evaporation of lead iodide (PbI$_2$) and methylammonium iodide (CH$_3$NH$_3$I / MAI) for the CH$_3$NH$_3$PbI$_3$ (MAPbI$_3$ / MAPI) perovskite shell. We present a detailed characterization of the nanostructures by (S)-TEM and XRD, revealing a different crystallization of the hybrid perovskite depending on the template: while the growth on H$_2$Pc nanowires induces the typical tetragonal structure of the MAPI perovskite, a low-dimensional phase (LDP) was observed on the one-dimensional TiO$_2$ nanotubes. Such a combination yields an unprecedentedly stable photoluminescence emission over 20 hours and over 300 hours after encapsulation in polymethyl methacrylate (PMMA) under different atmospheres including N$_2$, air, and high moisture levels. In addition, the unique one-dimensional morphology of the system, together with the high refractive index HP, allows for a strong waveguiding effect along the nanowire length.

## 1. Introduction

The last 15 years have witnessed an increasing interest in the development of halide perovskites (HP), such as the pure metal and hybrid organometal perovskites as photo-absorbers in third-generation solar cells with efficiencies reaching 26.1 % in single-junction and 33.9 % in tandem perovskite-silicon cells.[1] Even though issues such as long-term stability and reproducibility of the cell devices are yet to be solved, the photophysical properties of these materials make them impressively attractive for the development of optoelectronic systems.[2,3] Morphology, size, and structure of the HPs are critical in all these applications and essential to synthesizing low-dimensional systems in a highly reproducible way. In this context, low-dimensional perovskites such as 0D nanocrystals and nanoparticles, 1D nanowires, nanorods, and nanotubes, and 2D nanosheets have been recently reported.[4,5] Most of these examples showcase a a 3D-networked perovskite crystalline structure, different, from the 2D Ruddlesden-Popper perovskites.[2] 1D HP nanostructures have already been explored as lasers, photoabsorbers, and passivation fillers in solar cells, light-emitting diodes (LEDs), and photodetectors[2,3] due to their high crystallinity, superior charge carrier mobility, low defects, enhanced mechanical properties and surface area, tuneable bandgaps, anisotropic electrical/optical properties, and high photoluminescence quantum yield.[2,3] Direct on-substrate formation of these nanomaterials is also decisive to improve the level of integration in devices. Moreover, 1D perovskites possess a strong potential as a model system for a full elucidation of shape/size-dependent relationships and properties such as quantum confinement and charge transport.[6] Some examples in the literature have also demonstrated more robust stability than the counterparts of polycrystalline thin films, as the reduced lateral dimension would block the diffusion of water or oxygen.[6] In such an exciting context, different approaches have been proposed to fabricate low-dimensional nanomaterials, including solution-based, vapor transport, vapor phase synthesis, hot injection methods, templating, and combined solutions.[2,3] One of the major advantages of the use of template-assisted methods is the high definition and reproducibility of the shape of the nanostructures, particularly in the case of nanowires and nanorods. The most extended approach is the use of hard templates based on anodized aluminum oxide (AAO) and porous alumina membranes that have already been exploited for the fabrication of $MAPbI_3$, $MAPbBr_3$, $MASnI_3$, and $Cs_2SnI_6$.[2,7–11] However, an important drawback of this approach is the deleterious effect on the HP during the etching processes required for the removal of the template.

In this article, we present an alternative vacuum and plasma-based soft-template method for the preparation of core@shell nanowires/nanotubes formed by methylammonium lead iodide (MAPI) perovskite on one-dimensional supported nanostructures made of single crystalline metal-free phthalocyanine ($H_2Pc$) as a soft-template and nanoporous $TiO_2$. We have recently demonstrated the advantages of using this soft template approach for the formation of core@shell and core@multishell nanowires (NW) and nanotubes (NT) with different organic, metallic, and metal oxide compositions.[12–16] The process is carried out under mild vacuum and temperature conditions (below 350°C) and is fully compatible with processable substrates such as silicon, fused silica, glass, polymers (PET, PDMS), metal foils and meshes, and even cellulose.[16] To the best of our knowledge, this is the first time that this type of approach has been applied to the fabrication of 1D perovskite nanostructures. We will show below the fundamentals for the formation of $H_2Pc$@MAPI nanowires and $TiO_2$@MAPI nanotubes, the crystalline structure of the HP and their optical properties, focusing on the effect of the nanoscale arrangement on the photoluminescence and light-guiding properties. $TiO_2$@MAPI nanotubes exhibit outstanding photoluminescence stability even in a rich-moisture environment. We thoroughly compare the performance of the 1D nanostructures with polycrystalline thin films prepared under the same conditions also, including encapsulation with PMMA. The long-lasting stability is discussed under the premise of nanostructure and crystalline characteristics and the formation of a low dimensional perovskite (LDP) crystalline phase. Although some LDP halide perovskite crystalline phases have been reported with a higher stability such as the 2D Ruddlesden-Popper perovskites, the MAI-rich MAPI LDP phase reported here has not been yet related with an stability improvement.[17,18]

## 2. Results and Discussions

**Schematic 1** shows the multistep vacuum procedure applied to synthesize H$_2$Pc@MAPI NW and TiO$_2$@MAPI NT. Step 0) in the schematic corresponds to the growth of the seed layer (in this case, an amorphous, mesoporous, and nanocolumnar TiO$_2$ thin film) required for the crystalline growth of the 1D soft-templates. The first case for the synthesis of H$_2$Pc@MAPI NW (top side way of the schematic) consists mainly of two steps (see Experimental section for further details): 1-a) *growth of single-crystalline organic nanowires* (ONW) through self-assembly by pi-stacking of planar extended aromatic H$_2$Pc molecules by thermal evaporation. The approach for the growth of the ONW provides a reproducible and feasible route for the controlled formation of small-molecule NW (regarding density and mean length) supported on an ample variety of substrates.[16] The H$_2$Pc nanowires are highly flexible and grow along random directions with respect to the substrate, presenting a square or rectangular footprint and an extremely flat and pristine surface (see **Figure S1** in the Supporting Information section).[19–21] 2) *formation of the perovskite shell* by co-evaporation of the PbI$_2$ and CH$_3$NH$_3$I on top of the as-grown ONW. The procedure for the synthesis of HP crystals, in this case MAPI by co-evaporation of organic and inorganic counterparts, was previously reported by other authors in references [17,22,23]. In this study, we have adapted the deposition methodology to our system, allowing the formation of high-quality MAPI crystals with precise control in the shell thickness (and equivalent thin film) thickness.

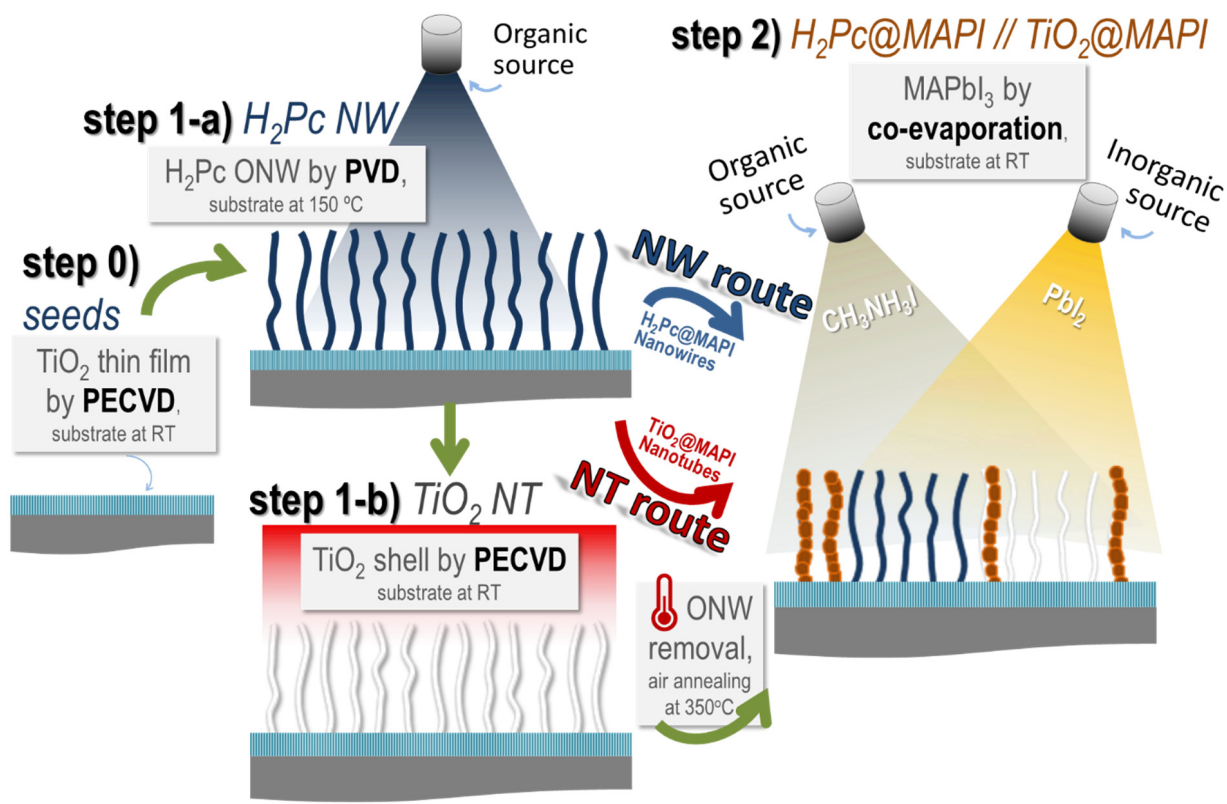

**Schematic 1. Step-by-step procedure for the fabrication of H₂Pc@MAPI nanowires and TiO₂@MAPI nanotubes.** Step 0) Thin layer deposition of TiO₂ by Plasma Enhanced Chemical Vapor Deposition (PECVD), which serves as seeds for the next step. Step 1-a) Growth of H₂Pc ONW by Physical Vapor Deposition (PVD). Step 1-b) TiO₂ thin film by PECVD deposited on top of the H₂Pc ONW, followed by an air annealing at 350ºC to remove the H₂Pc, thus remaining TiO₂ Nanotubes. The last step 2) consists of the co-evaporation of CH₃NH₃I and PbI₂ to fabricate MAPI in a core@shell structure: If step 2) is performed on the nanostructures obtained at step 1-a) or step 1-b), the H₂Pc@MAPI NW or TiO₂@MAPI NT are obtained, respectively.

The corresponding SEM images of the H₂Pc@MAPI NW at different magnifications are shown in **Figure 1** a, c, e) in planar view, as well as a cross-section in panel d) (in blue color). The images show how the perovskite is deposited along the ONW, producing, at first sight, a core@shell H₂Pc@MAPI 1D nanostructure without the presence of other MAPI nanostructures such as platelets or conglomerates. The mean length and diameter are about 4 μm and 150 nm, respectively (see the statistical analysis of the diameters in **Figure S2**). From **Figure 1** c), it can be noted that the length dispersion of the H₂Pc NW is broad, and the longest ones reach up to 10 μm and are bent, forming arcs randomly oriented. The thickness of the HP shell is homogeneous along the NW's length, with a slight decrease at the bottom near the interface with the substrate. This difference in thickness from the top to bottom apexes is the consequence of a self-shadowing effect intrinsic to the directionality of the vacuum deposition process. The higher magnification views (**Figure 1** e) show the distribution of the HP as a smooth and continuous shell. Surprisingly, a close inspection of the cross-sectional views of the NW reveals that the microstructure does not correspond to the expected coaxial architecture. The perovskite layer grows preferentially along one side of the ONW, as shown in **Figure S3**. This is likely due to the high directionality of the vapor evaporation technique and the low surface mobility and high sticking coefficient of the perovskite on the organic surface, resulting in a shell with an asymmetric section.

The physical vapor deposition method applied in step 1-a) (see **Schematic 1**) for the formation of the ONW is extendable to other molecules such as porphyrins and perylenes,[19,20] all of them deploying at least one absorption band in the visible, which can interfere with the optoelectronic applications of the MAPI.

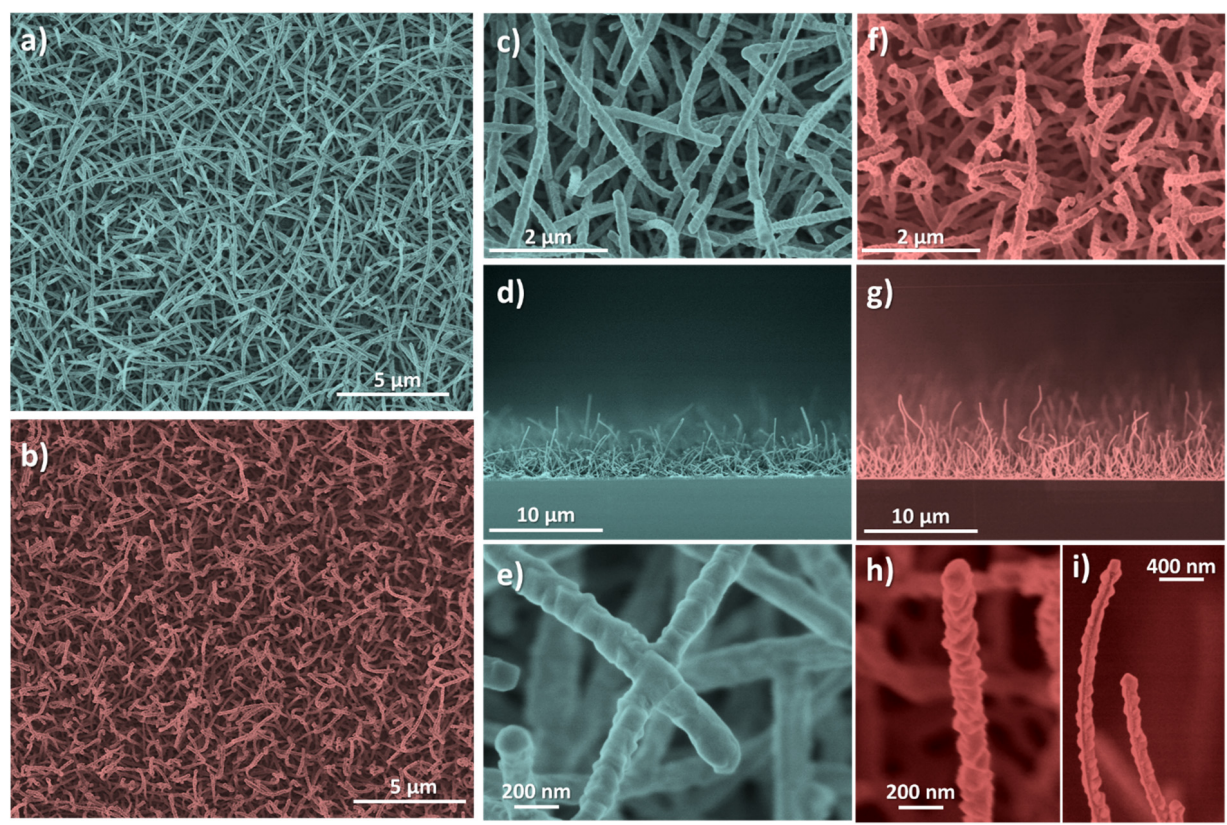

**Figure 1. Formation and microstructure of H$_2$Pc@MAPI nanowires and TiO$_2$@MAPI nanotubes.** a, c-e) SEM micrographs of the H$_2$Pc@MAPI NW (blue color): top-view (a,c,e) and cross-section (d). b, f-i) SEM micrographs of the TiO$_2$@MAPI NW (red color): top-view (b,f, h), cross-section (g,i).

To avoid interference of H$_2$Pc with MAPI absorption bands, we included a second case of study using a fully transparent 1D template. Thus, in step 1-b) (see **Schematic 1**), we generate a thin shell of nanoporous TiO$_2$ (nominal thickness ca. ~5 nm, see **Figure S1**) formed at room temperature by plasma-enhanced chemical vapor deposition (PECVD) on the ONW and yielding a full conformal coverage. As previously reported, the plasma sheath electric field provides a higher verticality to 1D nanostructures, as can be observed in Figure 1g) and S1 f).[12–14,24] The post-annealing of the samples in air at 350 °C ensures the complete removal of the H$_2$Pc organic core providing the formation of transparent TiO$_2$ NT. **Figure S4** gathers the evolution of the UV-vis transmittance spectra and the comparison between the blueish photograph characteristic of the H$_2$Pc NW sample and the fully transparent view of the TiO$_2$ NT sample. At first sight, the formation of the MAPI shell on the TiO$_2$ nanotubes (**Figure 1** b, f-i) presents high similarity with the H$_2$Pc@MAPI NW, but in high magnification (Figure 1 h-i), the appearance of a conglomerate of particles can be observed, in some cases with angular shapes. It is important to mention that the thickness of the multi-shell structure can be easily tuned by controlling the deposition time for both, TiO$_2$ and MAPI, thus, it is possible to fabricate extremely high aspect ratio nanowires as those presented in the micrographs in **Figure 1** g, i).

We have also analyzed the crystalline characteristics and microstructure of the H$_2$Pc@MAPI NW and TiO$_2$@MAPI NT (hereafter NW and NT) by XRD and TEM (**Figure 2**). XRD patterns in **Figure 2** a) show the comparison between the NW, NT, and the equivalent thin film (TF), i.e. the layer grown on a bare fused silica substrate and Si wafer with an equivalent MAPI thickness (~100 nm). It should be noted that both the 1D nanostructures (NW and NT) and the thin film were deposited simultaneously, and none of them show traces of the presence of PbI$_2$ (which maximum peak should appear at 12.7º). The comparison between both systems is not straightforward. On the one hand, NW and thin film share the characteristic peaks at 14.1°, 28.4°, and 43.2°, assigned to (110), (220), and (330) of the MAPI tetragonal crystalline phase. The thin film shows a higher surface texturization accounted for by the predominance of the (110) and (220) planes in contrast to the NW, which presents the expected diagram of a randomly distributed polycrystalline sample, with the presence of peaks related to a broader variety of planes. The high-resolution TEM image of H$_2$Pc@MAPI NW in **Figure 2** d) displays the interplanar distances of 2.25, 2.79, and 3.15 Å compatible with (224), (310), and (220) of the tetragonal phase of MAPI perovskite. The Digital Diffraction Pattern of the TEM image isshown in the inset of Figure 2 d, where it can be easy noted the presence of several crystals with different orientations. On the contrary, the XRD pattern corresponding to TiO$_2$@MAPI NT presents a principal double peak in the range between 11 and 12.2º which has been related by other authors with the formation of low-dimensional perovskites (LDP) in the crystalline meaning of the term.[17,18] The origin of such features at low diffraction angles is related to a higher interplanar distance due to intercalated MA$^+$ cations. References [17] and [18] discussed the presence of the LDP as a consequence of an excess of MAI during the deposition when the sublimation temperature settled for the MAI is too high (>155 ºC). However, in our case, the deposition conditions are the same for all the samples studied (please note that they have been deposited simultaneously). The High-Resolution TEM image of Figure 2e) demonstrates a much higher interplanar distance of 7.6 Å, in agreement with the XRD measurements. The DDP pattern of the image also shows this increased planar distance (note that the DDP scale is the same in Figures 3 d) and e)), with diffraction maxima close to the center.

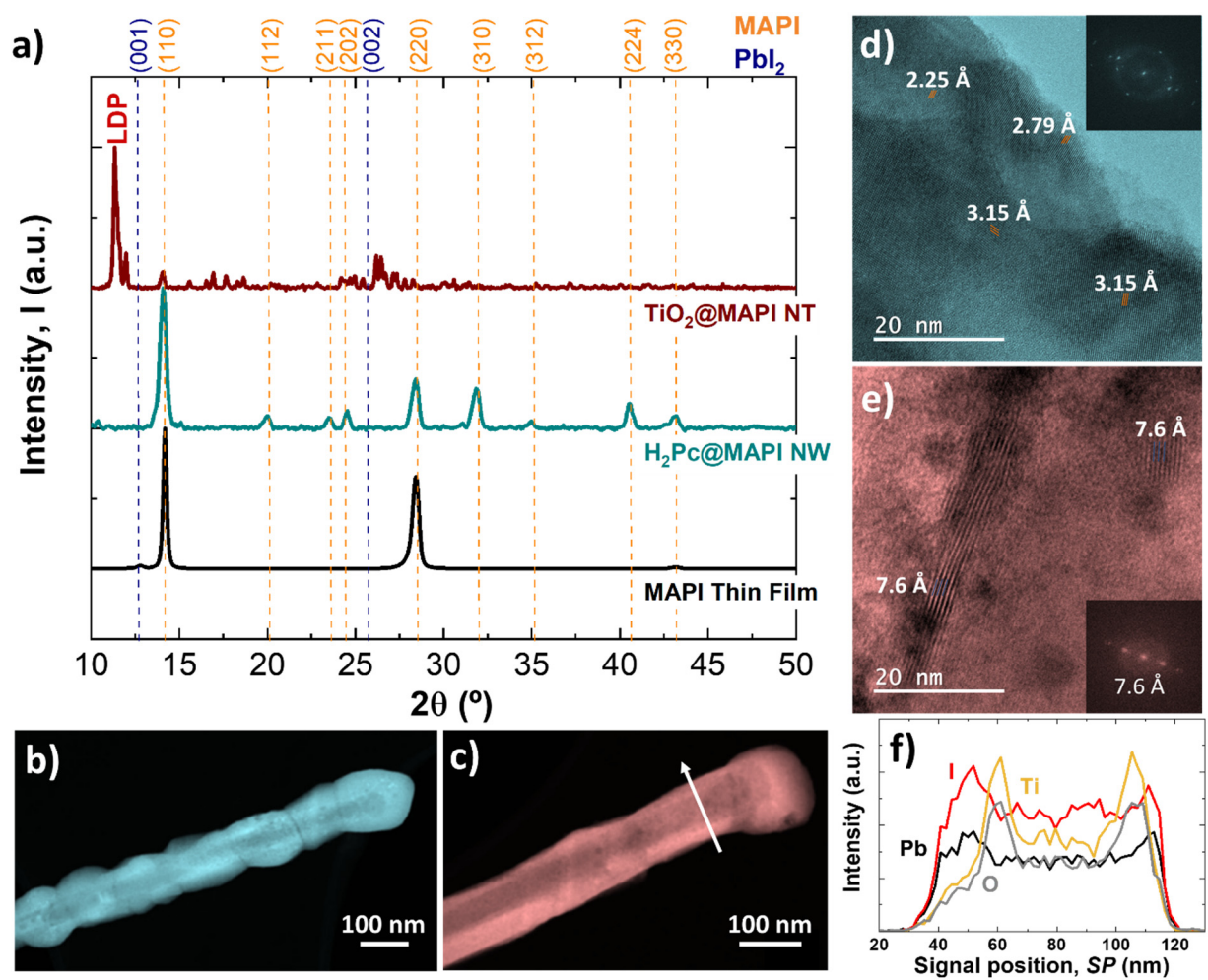

**Figure 2. Crystallinity and microstructure of core@shell H₂Pc@MAPI NW and TiO₂@MAPI NT.** a) XRD patterns comparing the growth of 1D perovskite-supported nanostructures with the thin film counterpart. b-c) HAADF-STEM images showing an individual H₂Pc@MAPI NW (b) and a TiO₂@MAPI NT (c). Panel f) shows the elements (Pb, I, Ti, O) distributions along the arrow highlighted in c) obtained by energy-dispersive X-ray spectroscopy for the TiO₂@MAPI NT. d-e) High-resolution TEM images of selected regions depicting lattice fringes compatible with the tetragonal and LDP crystalline phase for ONW@MAPI NW (d, and DDP (inset)) and TiO₂@MAPI NT (e, and DDP (inset)).

Furthermore, the appearance of LDP is usually accompanied by an optical absorption band at 370 nm related to the ground-state exciton from the zero-dimensional MAPI perovskite structure.[25] This feature is also present in the TiO₂@MAPI NT, as shown in **Figure 3 a)** (left, red curve highlighted with an arrow), corroborating the existence of LDP. The appearance of this band in the H₂Pc@MAPI NW cannot be easily observed, since the B-band of the H₂Pc appears at a very similar wavelength (see **Figure S4**). Other authors have analyzed the growth of MAPI on TiO₂ (110) single crystals and also related the enrichment of MAI to the hydroxyl groups on the oxide surface.[26,27] However, in our case, the appearance of these peaks in the XRD for the case of the NT is linked to the formation of the LDP phase with a much higher abundance (as deduced from the higher relative intensity with respect to the main tetragonal (110) perovskite peak), indicating that the LDP phase cannot be restricted only to the surface (the equivalent thickness is 100 nm). An additional study on MAPI growth on different thin films and 1D surfaces is shown in **Figure S5**. First, the growth on other $TiO_2$ surfaces such as thin films or the $H_2Pc@TiO_2$ core@shell structure (the latter without annealing at 350ºC to remove the organic phthalocyanine core) does not show this behavior, and only a minor LDP peak can be observed in the core@shell $H_2Pc@TiO_2@MAPI$ structure. This is a demonstration that the $TiO_2$ surface cannot directly induce the growth of the LDP phase. Second, the growth of MAPI on $H_2Pc$ TF (as well as on the $H_2Pc@MAPI$ NW) does not produce the appearance of this LDP, which diffractogram depicts a very similar pattern to the growth of MAPI on glass or $TiO_2$ TF, but with a small amount of $PbI_2$. Thus, it can be concluded that the appearance of this LDP phase is linked to both the hollow nanostructure and the $TiO_2$ surface chemistry. The first is evidenced by the fact that empty $TiO_2$ NT are required to grow the LDP as the main phase. We have corroborated that empty $TiO_2$ NT induce higher adsorption of the MAI molecules (when they are deposited alone without $PbI_2$), evident from the whitish appearance of the samples, which we speculate act as a deposit that supplies MAI during the perovskite deposition. The second evidence is also supported by the growth of the LDP MAPI phase (although with much lower intensity) in the $H_2Pc@TiO_2$ core@shell structure, which demonstrates that the effect of the $TiO_2$ surface chemistry cannot be ignored.

The crystallite sizes extracted from the XRD patterns show a smaller size for the 1D nanostructures, with mean crystallite sizes of 80, 50, and 30 nm for MAPI thin film, NT, and NW, respectively. Note that the crystalline size was calculated from the crystalline plane of tetragonal perovskite (110), thus having a lower accuracy for the NT because of its very weak intensity. These results are also consistent with the SEM images of **Figure 1**, where grains with angular shapes with sizes below 50 nm can be observed for both 1D nanostructures. The TEM characterization provides additional detailed information on the perovskite shell microstructure. In good agreement with the SEM results, **Figures 2** b-c) reveal a homogeneous distribution of perovskite crystals along the length and thickness of the 1D templates, which is also corroborated by the line-scan energy-dispersive X-ray spectroscopy analysis shown in **Figure 2** f) for the $TiO_2@MAPI$ NT (line scan performed at the location highlighted in c). These results demonstrate the homogeneous distribution of Pb and I of the MAPI in the outer shell of the nanostructure, while Ti and O are located in an inner shell (**Figure S6** shows the linescan profiles of both H₂Pc@MAPI NW and TiO₂@MAPI NT, for comparison). Such homogeneity in composition contrasts with the results reported in the literature for the synthesis of 1D HP nanostructures by catalytic vapor-solid (VS) and vapor-liquid-solid (VLS) approaches, usually presenting an accumulation of the metal particles at the top or at the bottom of the nanowires.[28]

A major advantage of our soft-template method is the direct fabrication of supported nanowires and nanotubes on processable substrates such as FTO and ITO/Glass[14,15] or flexible ITO/PET.[16] Here, we synthesized the NW and NT on a highly transparent fused silica substrate with a thin layer of TiO₂ working as a seed interface allowing for the on-support characterization of their optical properties. Thus, the optical (UV-vis spectra) and photoluminescence (PL) emission of the core@shell MAPI nanostructures are shown in **Figure 3** (see also **Figure S7** for the complete optical study) and have been analyzed in comparison with the thin-film counterpart. **Figure 3** a) displays the optical absorptance in the UV-vis for the ranges 350-600 nm (left) and 675-850 nm (right), this latter range also includes the normalized PL. The absorptance spectra of both 1D nanostructures (H₂Pc@MAPI NW and TiO₂@MAPI NT) and thin film show the typical bandgap onset at ~780 nm, proper of the MAPI perovskite.[29] Moreover, the Tauc-plots shown in **Figure S8** for the three types of samples show a similar band gap (1.62, 1.62, and 1.61 eV for thin film, H₂Pc@MAPI NW and TiO₂@MAPI NT, respectively) as expected for the bulk MAPI perovskite.[29] The fact that the bandgap is slightly lower for the TiO₂@MAPI NT could also be indicative of the presence of the LDP MAI-rich crystalline phase.

One important aspect extracted from the optical study is the enhanced absorptance of the 1D nanostructures. In the case of H₂Pc@MAPI NW, this effect can be related to the presence of the H₂Pc (which have strong and wide absorption bands, the B- and Q- bands at 370 and 590 nm, respectively, also shown in Figure S4). However, in the TiO₂@MAPI NT case, it can only be understood due to the presence of one-dimensional nanostructures. This significantly enhanced absorptance with respect to the MAPI thin film counterpart in the visible range (approximately 5 times higher in the 550-780 nm range), has been previously ascribed to light-trapping effects due to the nanostructuration of MAPI.[30]

The enhanced absorption of light has also an impact on the PL properties. The different PL intensities can be observed in the inset of **Figure 3** a) (also in the non-normalized PL spectra in **Figure S9**), which correspond to the IR photographs of the samples illuminated with a λ = 365 nm and collecting the images using a long-pass filter with a cut-on wavelength of 695 nm. It can be noted an enhanced PL emission of the TiO₂@MAPI NT with respect to the thin film for the same illumination intensity and camera acquisition parameters. However, we have corroborated a very similar PL quantum yield (PLQY) for the three samples of 0.012, 0.007, and 0.018 % for the thin film, NW and NT, respectively. This small difference between the PLQY indicates that the enhanced PL intensity is mainly due to the increased absorptance in the TiO$_2$@MAPI NT system. In the case of the H$_2$Pc@MAPI NW, the PLQY was even lower, which also suggests that a possible quenching effect can be occurring at the H$_2$Pc/MAPI interface (also supported by the lower PL emission intensity shown in **Figure S9**). However, the PLQY values are very low and thus are not fully reliable. In addition, the PL spectrum is slightly blue-shifted, from 769 nm for the thin film to 765 and 762 nm for the H$_2$Pc@MAPI NW and TiO$_2$@MAPI NT, respectively, which is in good agreement with the smaller particle size corresponding to the 1D nanostructures (with a mean crystallite size of 30 and 50 nm) as well as to the presence of LDP for the NT sample. We discard for both NT and NW any strong quantum confinement effect given the large mean thickness of both in comparison with the Bohr radius of MAPI (2.2 nm).[31]

When the samples were exposed to a 490 nm light source, the emission was evidently higher for the TiO$_2$@MAPI NT, followed by the MAPI thin films, and lastly, by the H$_2$Pc@MAPI NW. It is well-established that the presence of a scattering layer increases PL emission by enhancing light absorption, light trapping, waveguiding, and reducing reflection. In the case of TiO$_2$@MAPI NT, not only an enhanced PL emission is observed at perpendicular, but also at oblique angles. This effect is shown in **Figure S10**, where while the PL of a MAPI thin film at 25º drops to ca. 50% (with respect to the perpendicular PL), the TiO$_2$@MAPI NT maintain almost the same PL intensity, with a value above 99%. At more grazing angles this effect is even more pronounced, producing a drastic decrease of the PL of the thin film to 3%, while TiO$_2$@MAPI NT is above 40%. In the case of the H$_2$Pc@MAPI NW, the samples present a strong reduction of the photoluminescence emission. Two main factors can be accounted for such a feature, on one side, the absorption of light by the H$_2$Pc competing with the MAPI. On the other side, the quenching of the PL can be also related to efficient charge injection at the interface between the H$_2$Pc single crystalline nanowire and the MAPI. In this regard, numerous authors have previously reported on the benefits of the use of porphyrins and phthalocyanines as HTL and ETL.[32–35] Specifically, when dealing with evaporated HP, organic and polymeric interfaces have been demonstrated to outperform the use of inorganic ETLs such as TiO$_2$.[26,36]

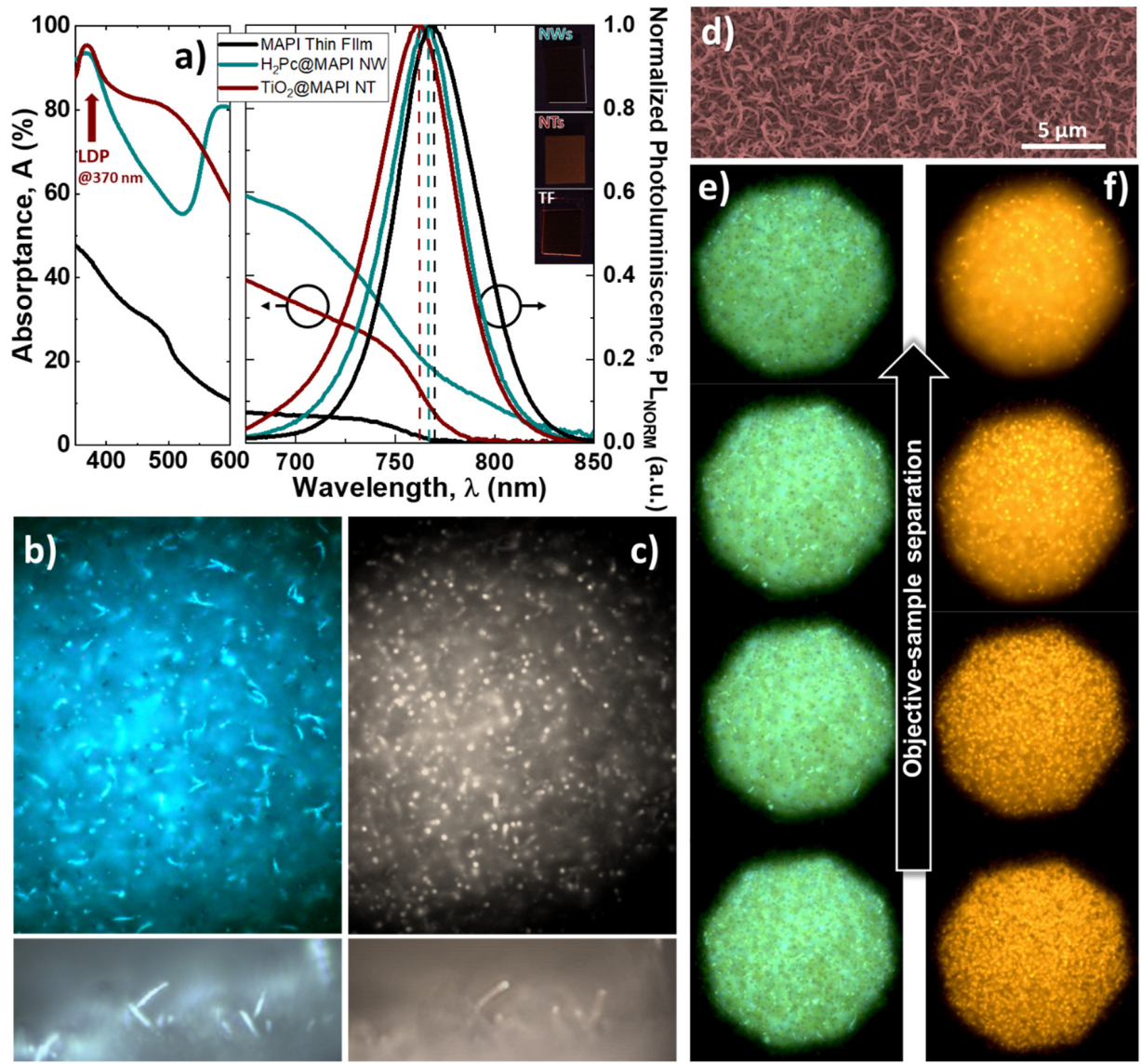

**Figure 3. Photoluminescence and waveguiding effect for core@shell MAPI nanowires and nanotubes.** a) UV-Vis absorptance and photoluminescence ($\lambda_{exc}$=490 nm) spectra comparing the systems of MAPI thin films, H$_2$Pc@MAPI NW and TiO$_2$@MAPI NT for equivalent thickness of 100 nm; the insets show characteristic PL photographs of the three surfaces illuminated with a $\lambda$=365 nm, and collecting the images using a longpass filter with a cut-on wavelength of 695 nm. Optical (b, e) and PL (c, f) ($\lambda_{EX}$=460-490 nm, $\lambda_{EM}$>550 nm) images were acquired for H$_2$Pc@MAPI NW (b,c) and TiO$_2$@MAPI NT (e,f), respectively. The sequence in e-f) are the corresponding microscope images at different sample-to-objective distances. d) Low magnification SEM top view image of the TiO$_2$@MAPI NT.

Long carrier lifetimes (fostered by reduced charge carrier trapping), long diffusion lengths, and low non-radiative recombination rates have made halide perovskite nanowires ideal materials for waveguiding and lasing.[37] Most of the examples of waveguiding effects in the literature develop around the exploitation of single crystalline HP nanowires, in which each NW serves as a waveguide and the two end facets form a Fabry-Perot cavity for optical amplification.[37–40] In **Figure 3** b – f), we explore the effect of the one-dimensional arrangement of the polycrystalline MAPI on such a nanophotonic property. **Figure 3** b-c) and e-f) compare the optical (b,e) and PL (c,f) images for $H_2Pc$@MAPI NW and $TiO_2$@MAPI NT, correspondently. The PL images show that the light is predominantly emitted from the tips of the NW and NT. Indeed, in the case of the $TiO_2$@MAPI NT, the vertical alignment of the nanostructures (see **Figures 3** d) and 1 g)) allows us to observe a plane-by-plane enhanced emission from the tips. From **Figure 1** g), it can be noted that although most of the tubes have a length in the 2-3 um range, there is a significant dispersion, finding vertical tubes as long as 8 μm. Thus, **Figure 3** e-f) showcases the light emitted by the NT of different lengths, in which every image corresponds to a z-step of ca. 2 μm (note that the 100x objective used with NA=0.9, has a depth of field of around 1 μm), with the top one corresponding to the highest separation (longer tubes). In a previous article, we reported on the waveguiding effect on perylene@$TiO_2$ and perylene@ZnO nanowires enhanced by the high contrast between the refractive index of the metal oxide shells and the luminescence core.[24] In **Figure 3**, the waveguiding effect is related to the high refractive index of the MAPI perovskite (n=2.5 @ 780 nm)[41] with respect to air, which gives a critical angle of 23.5°. Thus, the PL light arriving at the MAPI/air interface with angles above 23.5°, is trapped in the 1D nanostructure, and can only escape at the tips.

As it is well known, a critical drawback of using organometal halide perovskites as photoactive materials is their poor stability under room conditions (exposition to air, moisture, oxygen, and light). Specifically, the reaction between oxygen molecules and methylammonium cations in the presence of humidity produces the hydrolyzation of the MAPI back to the original $PbI_2$ precursor. Previous articles have reported on the higher stability of 1D perovskite nanostructures in comparison with their thin film counterparts.[42] Thus, for instance, Z. Fan *et al*. presented the template fabrication of $CH_3NH_3SnI_3$ NW on anodic aluminum oxide with 840 times slower PL decreasing than the planar thin film.[28] However, this significant improvement in stability was attributed to the effective blockage of the diffusion of water and oxygen molecules due to the inorganic template. *Fu et al.* showed that $FAPbI_3$ nanowires with diameters in the order of 10 nm present a significantly enhanced photostability with respect to higher-diameter wires or thin films.[43] Many authors have related this stability to the enhanced crystallinity of smaller clusters and reduced presence of grain boundaries.[44] Other authors have also considered the higher water contact angle of microstructured and functionalized HP arrangements as a positive factor to enhance stability.[45] With these antecedents in mind, we analyzed the PL stability of NW, NT, and TF (**Figure 4**). Under room air conditions (T=25ºC, RH=50-60%) and continuous illumination with monochromatic light at λ=490 nm and with a power of 0.35 mW/cm$^2$, the evolution of the PL emission follows a similar trend for the three samples but with very different time characteristic. Note that **Figure 4** a) left and middle curves are normalized to the maximum and initial PL values, respectively. Thus, the emission intensity increases at the beginning of the illumination to then decay to values lower than 20% of the initial (middle curves in **Figure 4** a) after only 1.2 hours for the thin film, 5.2 hours for the NW, but reaching more than 20 hours for the NT (note the logarithmic scale for the time axis).

The initial increase in the PL intensity induced by illumination has been previously reported. However, the explanation for such an increase is not unique and some authors assign it to a light-induced lattice expansion of the perovskite,[46] surface self-passivation of halide vacancies,[47] or surface chemical modifications in the presence of oxygen or moisture.[48] In our case, the increase is always present for the three samples studied, and the biggest change with respect to the original value occurs for the thin film (220%), and is slightly lower for the 1D MAPI nanostructures (185 and 175 % for NW and NT, respectively). Nevertheless, the time needed to reach this maximum is very different: in the thin film case, this time is extremely low and occurs after 15 minutes, while for the NW and NT, is 2.3 and 6 hours, respectively. These results clearly show an enhanced photo-stability for the 1D nanostructures, especially for the TiO$_2$@MAPI NT. Moreover, considering that the absorptance at 490 nm is much higher for the NT than for the reference thin film (77 vs 27%), this enhanced photostability is even higher since the total energy absorbed by the NT is almost three times higher than for the TF.

The PL pictures (taken in the same way as in the inset of **Figure 3** a), are shown in **Figure 4** a), right) before (As grown) and after 5 (for NW) and 20 h (for NT) of illumination time (after the PL evolution characterization). First, the horizontal elongated darker area visible in the NW and NT cases after 5 h and 20 h, respectively, show a full degradation of the perovskite luminescence. Although this elongated darker area is bigger for the reference thin film than for the NW and much bigger than the NT case, it is worth mentioning that the excitation monochromator slits have remained the same for all the PL evolution experiences. The reason for this bigger degraded area is also related to the enhanced stability of the 1D nanostructures: the illumination area presents a certain Gaussian profile in intensity, thus provoking that close to the edges even if the illumination intensity is much lower, is sufficient to induce the full degradation of the reference MAPI thin film. By contrast, for the NT sample, the lower intensity at the edges of the illumination area has not been enough to degrade the nanostructures, but to enhance the PL. This is visible with an increased PL intensity at the borders of the illuminated area for the NT, and to a lesser extent in the NW samples. In fact, the XRD analysis of the TiO$_2$@MAPI NT system after several weeks stored in the lab at a relative humidity of ca. 40-60 % only suffers modifications in the relative intensity between the LDP and (110) tetragonal perovskite peaks, without the presence of the PbI$_2$ characteristic peaks (see **Figure S11**). This is a very striking result, since the aging only produces the transformation of the LDP to the MAPI tetragonal structure, without producing any degradation to the PbI$_2$, as it happens, for example, in the thin film aged for a similar time at the same storage conditions (See **Figure S11**). This feature also demonstrates the higher stability of the TiO$_2$@MAPI NT concerning the thin film counterpart.

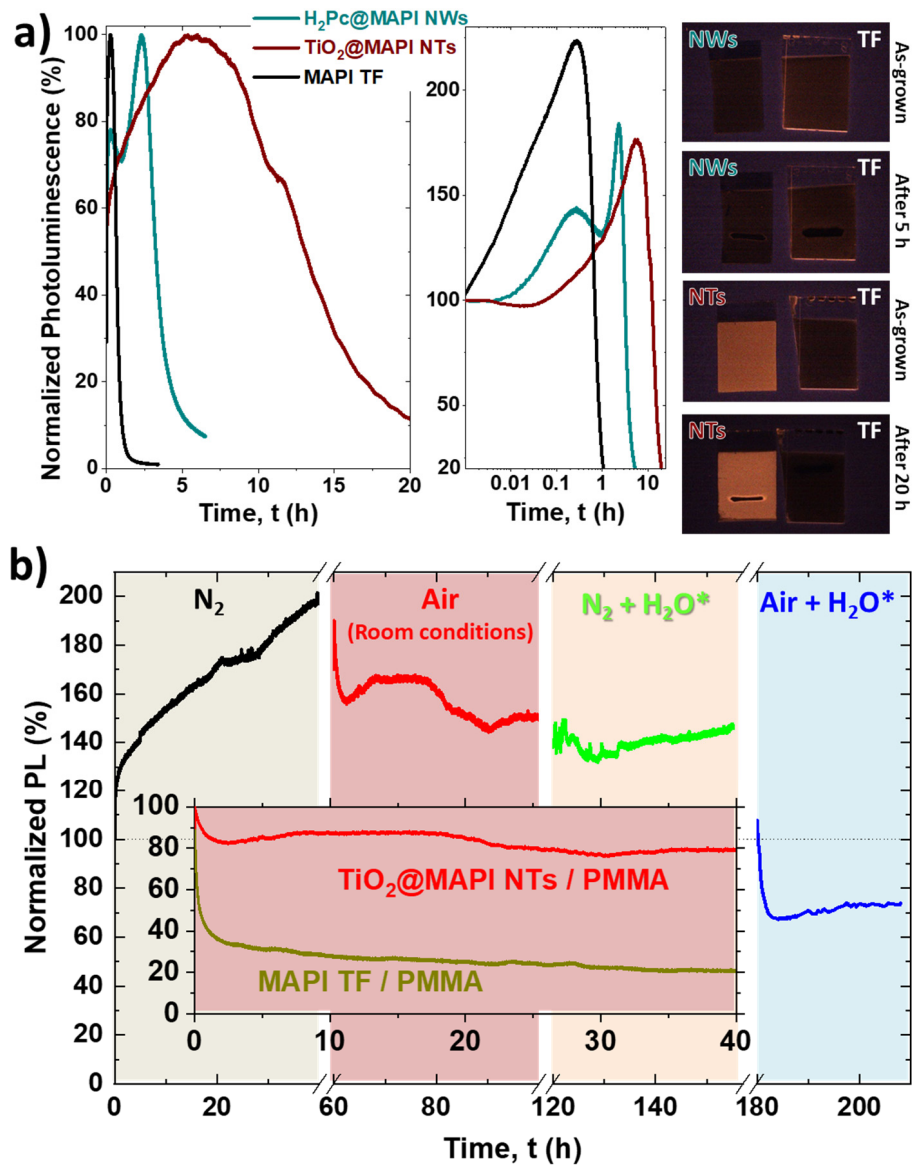

**Figure 4. Photostability of MAPI thin film, nanowires, and nanotubes.** a) *Left and middle.* Stability test of 1D MAPI nanostructures and thin films in air atmosphere and illuminated with monochromatic light ($\lambda_{EX}$=490 nm, with a power of 0.35 mW/cm$^2$), monitoring the intensity at the maximum emission wavelength. The time evolution curves have been normalized to the maximum (left) and to the initial (middle) PL intensity value. Note that the time axis in the middle graph is in logarithmic scale. *Right.* IR photographs of the samples before (as grown) and after the stability test comparing nanostructured samples with equivalent thin films. b) Hard stressing stability test of TiO2@MAPI NT sample encapsulated in PMMA illuminated with monochromatic light ($\lambda_{EX}$=490 nm), at intervals of 40 h sequentially exposed to N$_2$, air (room conditions), N$_2$+100% RH, and air+100% RH atmospheres (24h relaxing period in inert atmosphere between cycles). The inset shows the fast decay on PL of the reference MAPI thin film exposed to the lab environment in comparison with the nanostructured sample (both encapsulated in PMMA).

The encapsulation of hybrid metal halide perovskite solar cells and other optoelectronic devices with different polymeric solutions has been previously demonstrated to improve the stability of these devices under room environmental conditions.[49] The encapsulation diminishes the interaction with the atmosphere, and allows to separate the intrinsic stability of the perovskite with respect to the chemical reactions (which are usually accelerated by the continuous illumination) of the MAPI with the moisture and/or oxygen. To study the enhanced intrinsic stability of the $TiO_2$@MAPI NT, the sample was encapsulated in PMMA (see Experimental section). **Figure 4** b) shows the PL evolution during intervals of 40 hours of continuous illumination under different atmospheres, i.e., i) pure $N_2$, ii) air (room conditions), iii) moisture-saturated $N_2$ (using a water bubbler), and iv) moisture saturated air. In the first 40 h of the experiment under an $N_2$ atmosphere, it can be noted a continuous increase of the PL with the illumination time, which reaches approximately 200% of the initial value. The exposure to air room conditions produces a slight decrease and later stabilization of the PL, which is again constant for the next experiment under a nitrogen atmosphere saturated with water. Note that the PL stability under air is significantly enhanced with respect to the previous experience shown in **Figure 4** a), in which after 20 hours of continuous illumination, the unencapsulated sample reduced its PL to 20 % of the initial value. In this case, the PMMA encapsulated NT do not present any degradation, an appealing result considering that the sample has been illuminated during 120 h in three different atmospheres, one of them containing a high level of moisture. The last environmental test is the most aggressive one, a moisture-saturated air atmosphere under continuous illumination, which provokes an initial decrease of the PL to around 65 %, to then slightly increase to finally stabilize to 75%.

It is worth noting that even if the PMMA encapsulation does not completely block the effect of the atmosphere (it can also be understood with the SEM images, of the encapsulated samples shown in **Figure S12**, since the PMMA does not cover completely the longest NT), it has been greatly inhibited since after 120 h of illumination with different atmospheres, the only experience that produces an apparent decrease in the PL is the most aggressive one, air plus moisture. But even in this case, the decrease is not very drastic, since it only experiences a decrease from 110 to 75 % during 30 hours of illumination. In comparison with the unencapsulated $TiO_2$ NT, which completely degrades the PL to values below 20% in the first 20 h, this decrease is not very significant. This is particularly notable when considering the illumination history of the sample plus the aggressive atmosphere used. It is worth mentioning that for the case of encapsulated samples, the PL of the nanotubes also presents a much higher intrinsic stability than the corresponding to the equivalent thin film emission, as shown in the inset of **Figure 4** b). Thus, while the NT decreased to 80 % of the initial value of the experience, the thin film was quenched down to 20 %, both after 40 hours of light soaking. Such a successful result led us to carry out a long-term stability experiment covering more than 7 months under dark storage in laboratory conditions. **Figure S13** shows the evolution of the PL intensity emission during the period, presenting a decrease to 35 % of the original emission. However, it needs to be remarked that the dark storage evolution measurements suffer a very high dispersion in the results since the final PL intensity depends on many variables such as the illuminated area, the intensity of the lamp, or the collection angle, which can be varied after such prolonged period. In any case, these PL measurements can assure that the samples are stable when encapsulated in PMMA, although a certain degree of structural or chemical modification cannot be discarded. As mentioned before, this enhanced stability was corroborated by XRD as shown in **Figure S11,** with an equivalent sample stored in the dark over 2 months, which showed a decreased intensity of the LDP-associated peak, but without visible degradation, since $PbI_2$ species are not present in the pattern. This enhanced stability, which is ascribed to the molecular arrangement of the ions in the LDP crystalline structure rather than to its morphological features, has been previously reported by other authors. Every octahedral $[PbI_6]^{4-}$ in a LDP phase does not share the 6 $I^-$ atoms with the next octahedra. This "shoulder to shoulder" arrangement of the octahedra improves the skeletal strength of the perovskite lattice. At the same time, the $[PbI_6]^{4-}$ are wrapped by the $MA^+$ cations, thus protecting the structure from moisture and oxygen, and providing the LDP with significantly enhanced stability compared to the 3D perovskite.[50,51] Although many efforts have been made in the past to synthesize purely low dimensional perovskites (from a molecular point of view) by tuning the organic cation separators (like for example in the 2D Ruddlesden–Popper perovskites), very few examples can be found in the literature about the synthesis and thus the properties of the MAI-rich MAPI perovskites reported here, which can open a new way to enhance the stability of hybrid halide perovskites.

## 3. Conclusions

We have demonstrated a reliable full vacuum/plasma methodology based on the use of supported $H_2Pc$ ONW as templates for the fabrication of 1D metal halide perovskites. The properties of the MAPI perovskite co-evaporated layers are highly dependent on the surface nanostructure. While the growth on flat substrates such as glass or Si resulted in a highly textured MAPI tetragonal crystalline structure (along the plane (110)), the deposition on $H_2Pc$ NW produced the same crystal perovskite phase with a much higher polycrystalline character.

By contrast, the deposition on TiO$_2$ NT induces a different crystalline growth, mainly dominated by the presence of an MAI-rich perovskite phase referred to in the literature as "Low Dimensional Perovskite" or LDP. Moreover, the existence of the LDP phase is corroborated not only by a clear peak in the XRD spectrum, but by the high resolution TEM images and by the presence of an additional optical absorption band at 370 nm, which has been reported by other authors.

The 1D MAPI nanostructures present an enhanced absorptance of light, which is from 4 to 6 times higher than the MAPI thin films reference in the 550-750 nm range. Although in the case of H$_2$Pc@MAPI NW, this enhanced absorptance can be due to the phthalocyanine absorption, in the TiO$_2$@MAPI NT can only be ascribed to light-trapping effects due to the nanostructuration of MAPI. As a result of this enhanced absorptance, the PL emission is significantly increased for the TiO$_2$@MAPI NT. In addition, the 1D nanostructures present an important waveguiding effect related to the high refractive index of the MAPI perovskite compared to air, producing the total internal reflection of the PL emission and confining the light into the perovskite which can only escape at the tips of the NW and NT.

The stability of the 1D MAPI nanostructures is significantly enhanced as demonstrated by their PL emission under prolonged light illumination periods. This PL evolution behavior, together with the increased absorptance of the 1D nanostructures, especially for the TiO$_2$@MAPI NT, demonstrates a 50 times higher stability. A very remarkable result observed by XRD is that the dark storage under laboratory conditions of TiO$_2$@MAPI NT for several months does not produce the degradation of the perovskite phase to PbI$_2$ as it happens for the MAPI thin film, but reduces the relative intensity of the LDP to (110) tetragonal perovskite phase. Other authors have ascribed this enhanced stability to the molecular arrangement of the ions in the LDP crystalline structure, which improves the skeletal strength of the perovskite lattice and the protection of the organic cations by the metal halide octahedra. The stability can be further increased for real photonic applications by encapsulating the samples in PMMA, which decreases further the interaction with the environment. This provides a very high stability for TiO$_2$@MAPI under aggressive environments.

Our template method provides different advantages for the straightforward implementation of 1D perovskites as advanced nanoarchitectures for optoelectronic applications since it is compatible with an ample variety of materials, including TCOs or preformed devices, and it is environmentally friendly since it uses mild vacuum and temperature conditions and does not require harmful solvents or produce toxic wastes. Our approach is straightforwardly extendable to other hybrid and inorganic HP processable by vacuum thermal deposition and compatible with compact, nanoporous, amorphous, or highly texturized shells of metal, metal oxide, or polymer-like compositions prepared by vacuum and plasma-assisted methods. In addition, the results reported here about the synthesis and properties of this LDP phase in TiO$_2$@MAPI NT open a new pathway to enhance the stability of hybrid halide perovskite for optoelectronic applications.

## 4. Experimental Section/Methods

*Synthesis of H$_2$Pc nanowires and TiO$_2$ nanotubes* (Step 0, 1-a, and 1-b, in Scheme 1). The fabrication of the 1D nanostructures was carried out by using a full vacuum and plasma soft-template method reported previously by our research group.[12–14,24] The supported 1D nanostructures were synthesized on Si (100) wafers and fused silica pieces. First, a "seed" layer of TiO$_2$ (~100nm) was deposited at room temperature by plasma-enhanced chemical vapor deposition (PECVD) using as organometallic liquid precursor titanium (IV) isopropoxide (TTiP) from Sigma Aldrich (step 0 in Scheme 1).[52] Then, it was grown H$_2$Pc (29H,31H-Phthalocyanine, β-form, 98%, Sigma-Aldrich) organic nanowires supported on top of the film according to the conditions of the method (step 1-a).[19] For the TiO$_2$ nanotubes synthesis (step 1-b), additional steps were required: 1. *TiO$_2$ shell*: H$_2$Pc ONW were covered conformally by a 5 nm TiO$_2$ shell using PECVD (similar condition of "seed" layer deposition); 2. *Organic core removal*: samples were annealed at 350$^O$C (+2$^O$C/min ramp, 2 h constant temperature, -2 $^O$C/min ramp till room temperature) in an air atmosphere for removing the H$_2$Pc core, leading to a 1D TiO$_2$ nanostructure fully empty.

*CH$_3$NH$_3$PbI$_3$ co-evaporation* (Step 2). Lead Iodide (PbI$_2$) and methylammonium iodide (MAI) were acquired from Sigma Aldrich and TCI (respectively) and used after preheating under high vacuum (5·10$^{-6}$mbar) during 1h at 200 $^O$C for PbI$_2$ powder and 100 $^O$C for MAI powder. Two independent Knudsen cells were employed for the sublimation of the precursors, positioned 15 cm away from the samples (kept at room temperature the whole process). The chamber pressure was fixed at a certain value (as was previously reported by other authors in references [23,25]), which in this study and for our chamber configuration was 1.0·10$^{-4}$ mbar. The process started establishing the deposition rate (R) at 0.2 Å/s for the PbI$_2$ source controlled using a quartz crystal microbalance (QCM) positioned at the same height as the samples. Parallelly, a second QCM positioned near the PbI$_2$ source serves as deposition rate control for this only precursor. After that, MAI powder was heated up progressively, controlling pressure increases, till reaching a stable temperature of 150 $^O$C.[17] MAI deposition rate could be difficult to control individually, as it is reported in the literature.[36,53,54] For this reason, it was set a specific chamber pressure and sublimation temperature to reach the correct stoichiometry.[17,23] During the co-deposition, the main QCM (near the samples) shows an increase in the deposition rate due to the presence of MAI. It is reported that the interaction of MAI with $PbI_2$ ascribes an increment in the sticking coefficient of the organometal compound.[53] Thus, this ΔR can be used as a measurement of the MAI deposition rate, comparing the values in the main and $PbI_2$ QCMs.[54] In this study, we have established this ΔR=0.06 Å/s as the optimum value for our chamber configuration. As a result, co-evaporated H₂Pc@MAPI NW, $TiO_2$@MAPI NT, and MAPI equivalent thin film samples were developed. The samples were transferred directly from the vacuum chamber to a $N_2$ dry box without exposure to air.

*PMMA encapsulation*. Samples were embedded in a polymethyl methacrylate (PMMA) polymeric matrix to test the stability against different stressing conditions. PMMA (Sigma Aldrich) was dissolved at 5%wt. in Chlorobenzene, CB, (Sigma Aldrich) under vigorous stirring for 12 h. The solution was spin-coated on top of samples in two steps: 2000 rpm during 10 s (step 1) and 5000 rpm during 30 s (step 2), and then dried at 70°C in a hot plate for 30 min. The whole process was developed into a $N_2$ dry box to avoid the exposure of perovskite samples to air.

*Morphological, structural, and chemical characterization*. SEM micrographs were acquired in a Hitachi S4800 working at 2 kV. The 1D nanostructures were dispersed onto Holey carbon on Cu or Ni grids from Agar scientific for TEM characterization. HAADF-STEM, HRTEM, and EDX line profiles were carried out with a FEI Tecnai G2F30 working at 300 kV and with a JEOL2100plus working at 200 kV, using low doses to avoid damage and structural changes. The crystal structure was analyzed by XRD in a Panalytical X'PERT PRO spectrometer operated in the θ - 2θ configuration and using the Cu $K_α$ (1.5418 Å) radiation as an excitation source. The crystallite size was determined with PANalytical X'Pert HighScore Plus software, which employs the Scherrer equation for the calculations.

*Optical characterization*. Transmittance, reflectance, and absorptance measurements were carried out in a PerkinElmer Lambda 750 UV/vis/NIR equipped with an integrating sphere. Photoluminescence spectra were acquired in a Jobin Yvon Fluorolog-3 spectrofluorometer using the front face configuration. Optical and photoluminescence micrographs were obtained using an Olympus BX51 microscope equipped with Lumenera's INFINITY3-3UR camera. IR images were taken from a Genie Nano-C2420 camera with a MidOpt LP695-27 near infrared longpass filter.

*Stability tests*. Photoluminescence of the samples was tested at different environments under constant illumination for established periods using a Jobin Yvon Fluorolog-3 spectrofluorometer. To control the atmosphere, an ad-hoc chamber was built andprovided with a gas inlet/outlet system, moisture-meter, and sealed fused silica illumination window (see **Figure S14**). Stability tests were carried out at constant illumination conditions ($\lambda$ = 490 nm, P = 0.35 mW/cm$^2$) for 40 h each. Sequentially, the samples were tested under different atmospheres in: (1) $N_2$, (2) room air conditions (RH=40-60%), (3) $N_2$+$H_2O$ (RH=100%), and (4) air+$H_2O$ (RH=100%).

**Supporting Information**

Supporting Information is available.

**Acknowledgments**

We thank the projects PID2022-143120OB-I00, PID2019-109603RA-I00, TED2021-130916B-I00 funded by MCIN/AEI/10.13039/501100011033 and by "ERDF (FEDER) A way of making Europe, Fondos Nextgeneration EU and Plan de Recuperación, Transformación y Resiliencia". The authors also want to thank CSIC for its financial support through the Intramural Project PIE 202260I156. The project leading to this article has received funding from the EU H2020 program under grant agreement 851929 (ERC Starting Grant 3DScavengers).

Received: ((will be filled in by the editorial staff))
Revised: ((will be filled in by the editorial staff))
Published online: ((will be filled in by the editorial staff))

Supporting Information

**Highly stable photoluminescence in vacuum-processed perovskite-based core-shell nanostructures**

*Javier Castillo-Seoane, Lidia Contreras-Bernal,\* T. Cristina Rojas, Juan P. Espinós, Andrés-Felipe Castro-Méndez, Juan P. Correa-Baena, Angel Barranco, Juan R. Sanchez-Valencia,\* and Ana Borras*

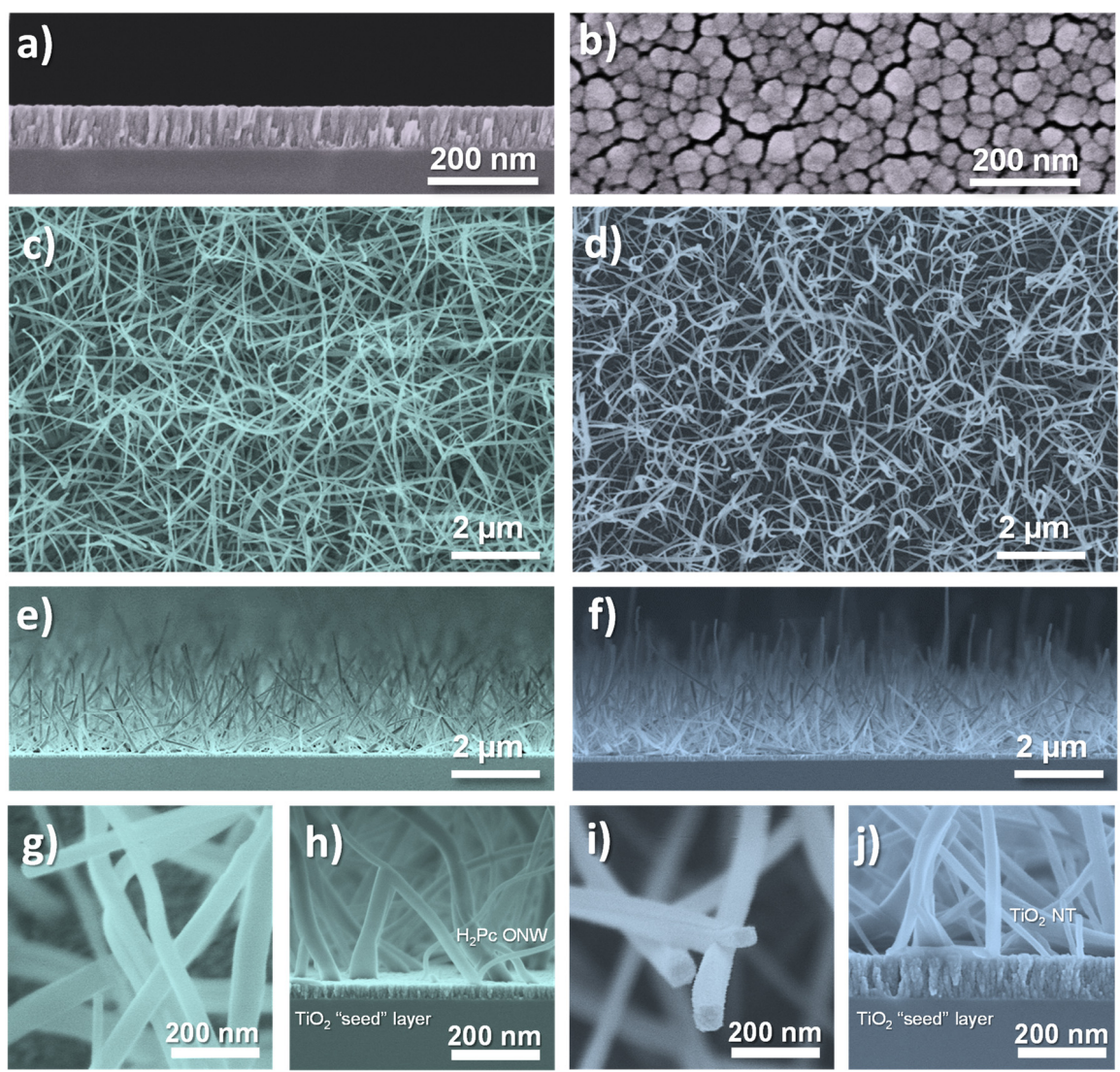

**Figure S1.** Representative SEM images of the $TiO_2$ "seed" layer (a, b), supported $H_2Pc$ ONW (c, e, g, h) and $TiO_2$ nanotubes (d, f, i, j) used as 1D templates.

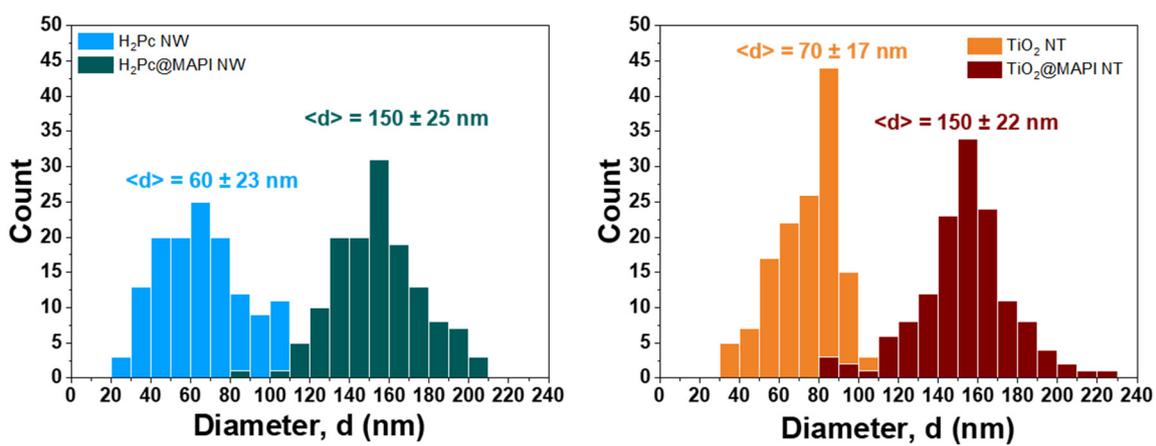

**Figure S2.** Statistical image analysis of 1D nanostructures diameters.

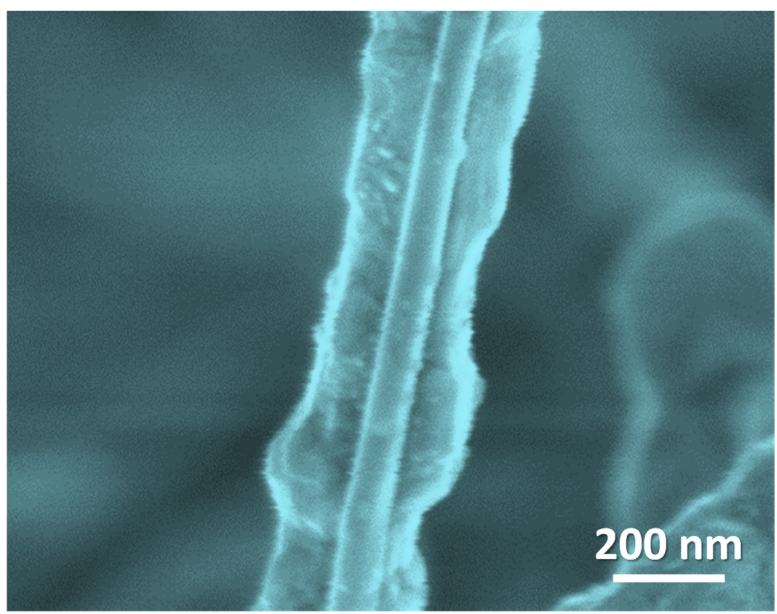

**Figure S3.** Cross-sectional SEM view at high magnification of the H$_2$Pc@MAPI nanowires.

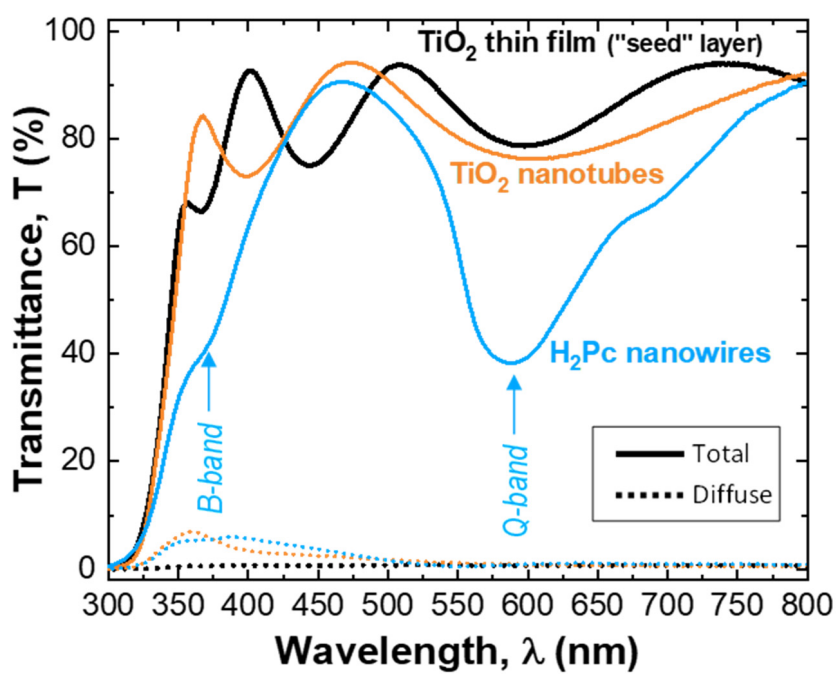

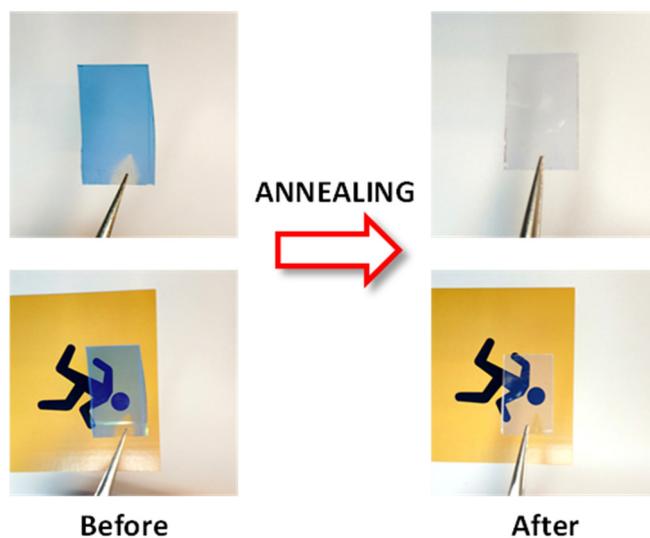

**Figure S4.** Top) UV-Vis transmittance spectra comparing $TiO_2$ thin film of equivalent thickness to the $TiO_2$ shell (thin film prepared during the same experiment on a flat fused silica substrate, refractive index at 550 nm = 1.89), sample corresponding to high density of $H_2Pc$ NW and the same sample after deposition of the $TiO_2$ shell and annealing. Bottom) Photographs of the sample before and after annealing at 350 ºC in air. The sample after annealing depicts a light whitish aspect due to the light scattering effect expected for the arrangement of micrometric length and randomly oriented nanotubes.

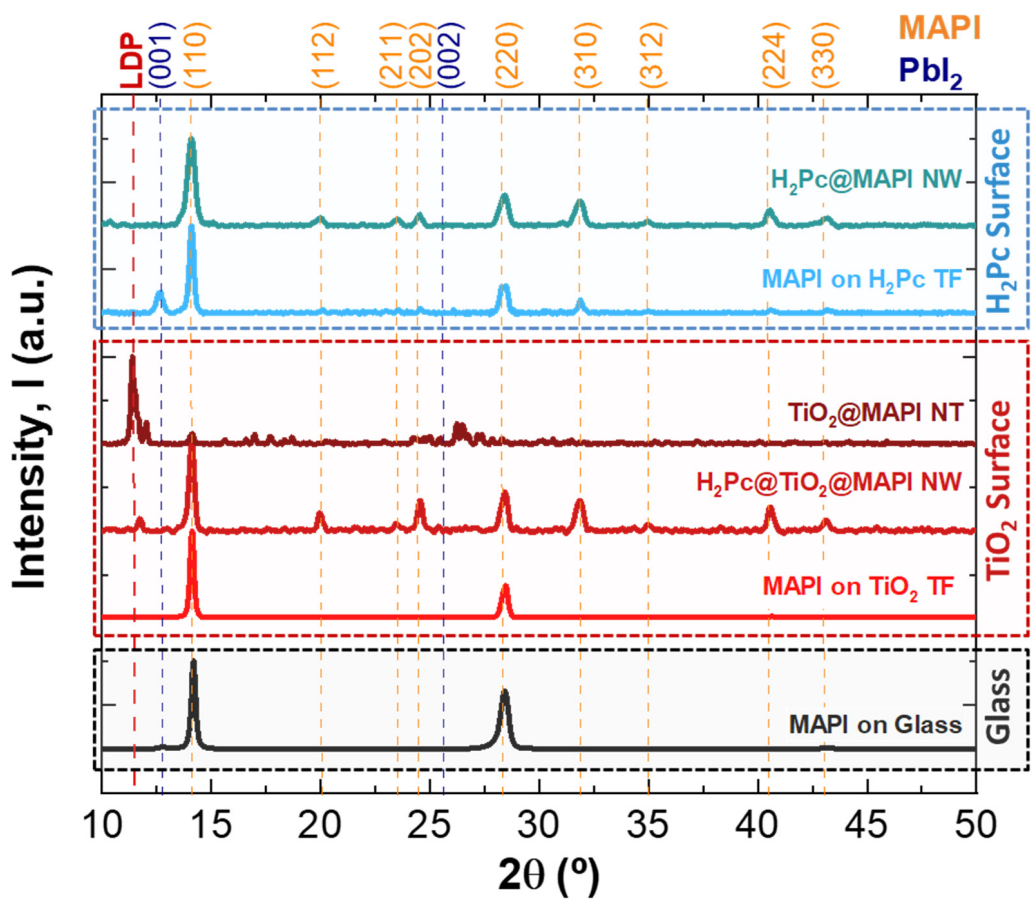

**Figure S5.** X-ray diffraction patterns (XRD) comparing the growth of MAPI on different surfaces, glass, TiO2 and H2Pc, and nanostructures as labeled. Vertical dashed lines are included to indicate the position of the peaks corresponding to PbI2 (blue), tetragonal MAPI (orange), and the LDP phase (red).

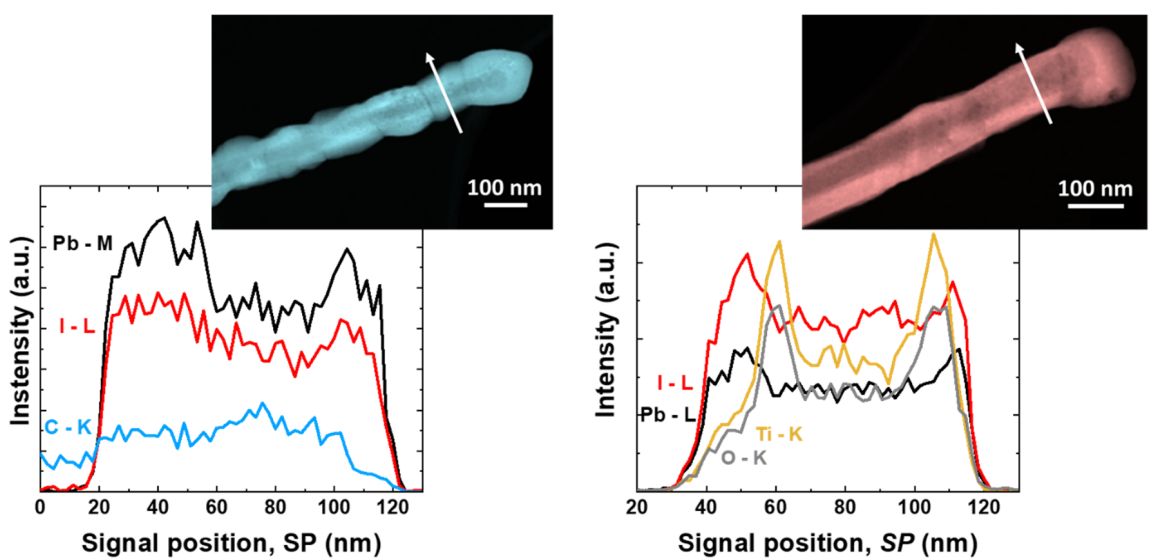

**Figure S6.** Line-scan Energy Dispersive X-ray Spectroscopy analyses of the cross-section profile of an ONW@MAPI NW (a) and a TiO2@MAPI nanotube (b).

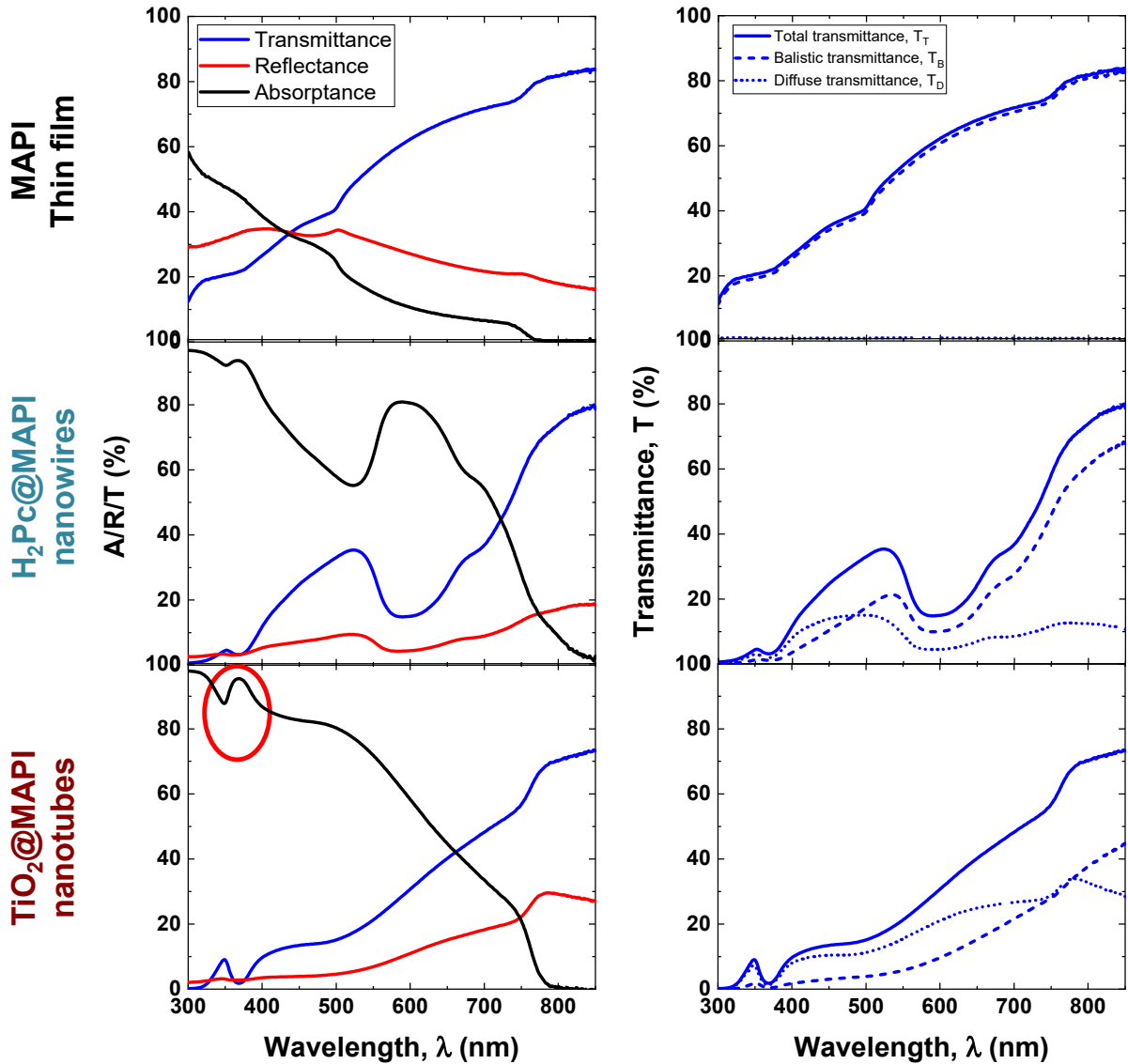

**Figure S7.** Complete optical study. *Left*) Reflectance (R, red), Transmittance (T, Blue), and absorptance (A, black curves). *Right*) Ballistic, Diffuse, and Total Transmittance curves. All the spectra have been measured for thin films (top), H$_2$Pc@MAPI NW (middle), and TiO$_2$@MAPI NT (bottom).

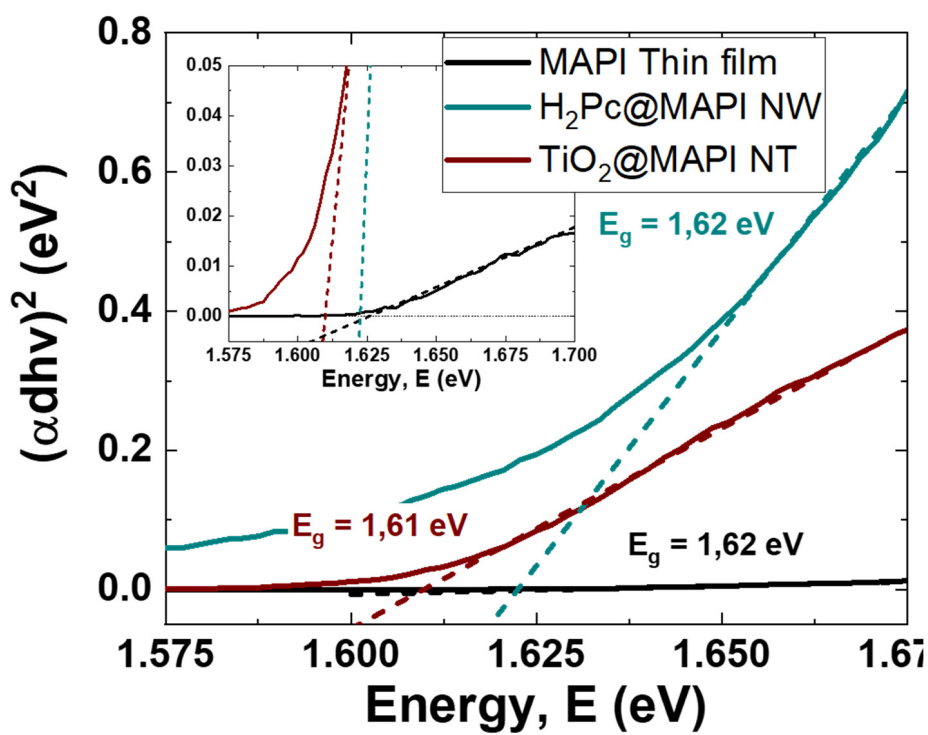

**Figure S8.** Tauc plots for the H₂Pc@MAPI NW (blue), TiO₂@MAPI NT (red), and MAPI thin film (black) deduced from the absorptance spectra shown in Figure S7

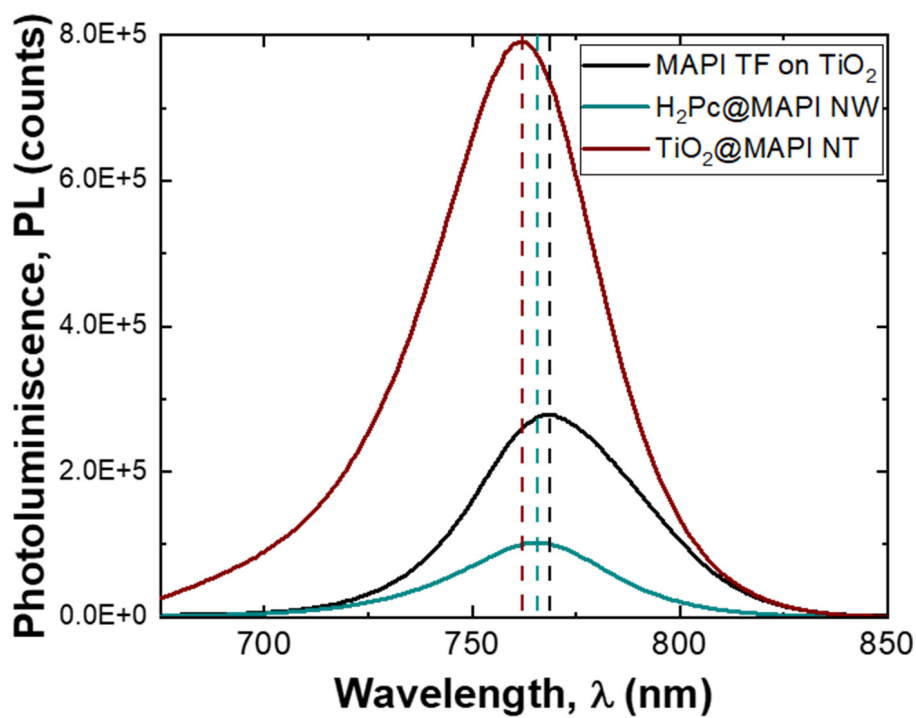

**Figure S9.** PL spectra shown in Figure 3 without normalization

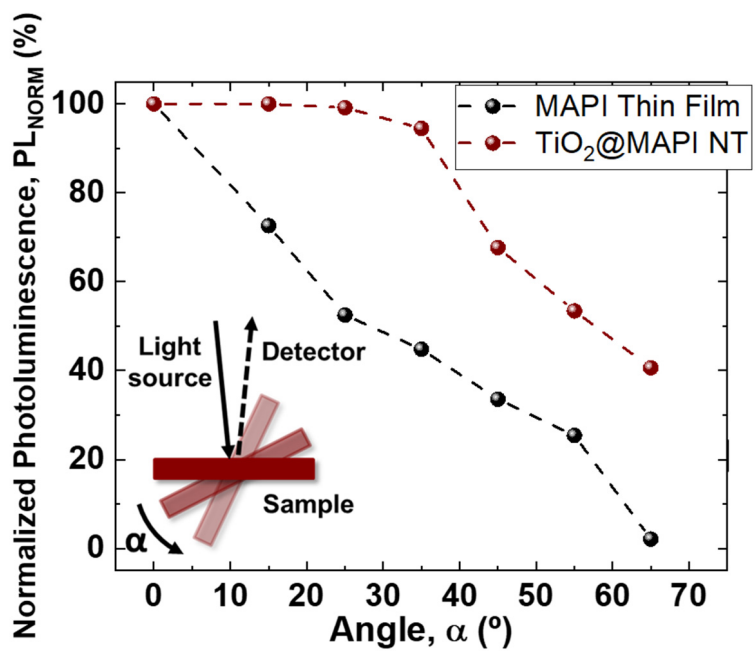

**Figure S10.** Normalized Photoluminescence of MAPI thin film (black) and TiO$_2$@MAPI NT (red) as a function of the emission angle.

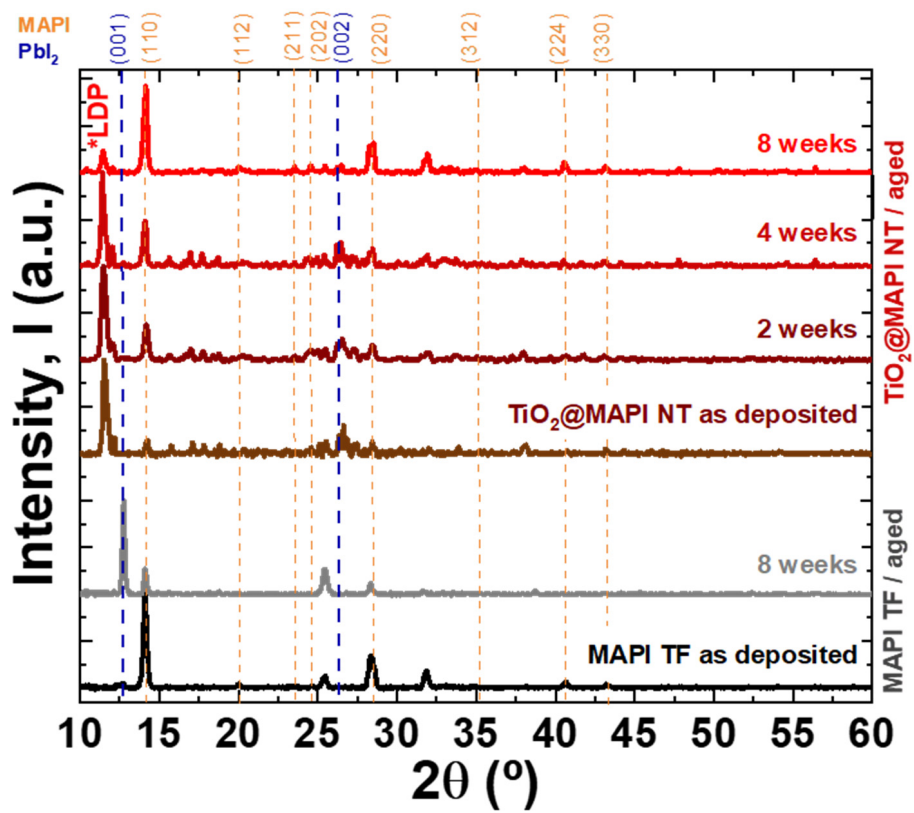

**Figure S11.** Comparison of the XRD patterns of the TiO$_2$@MAPI NT as-grown and after storing under room conditions for 2, 4, and 8 weeks. The MAPI thin film degradation after 8 weeks has been added for comparison.

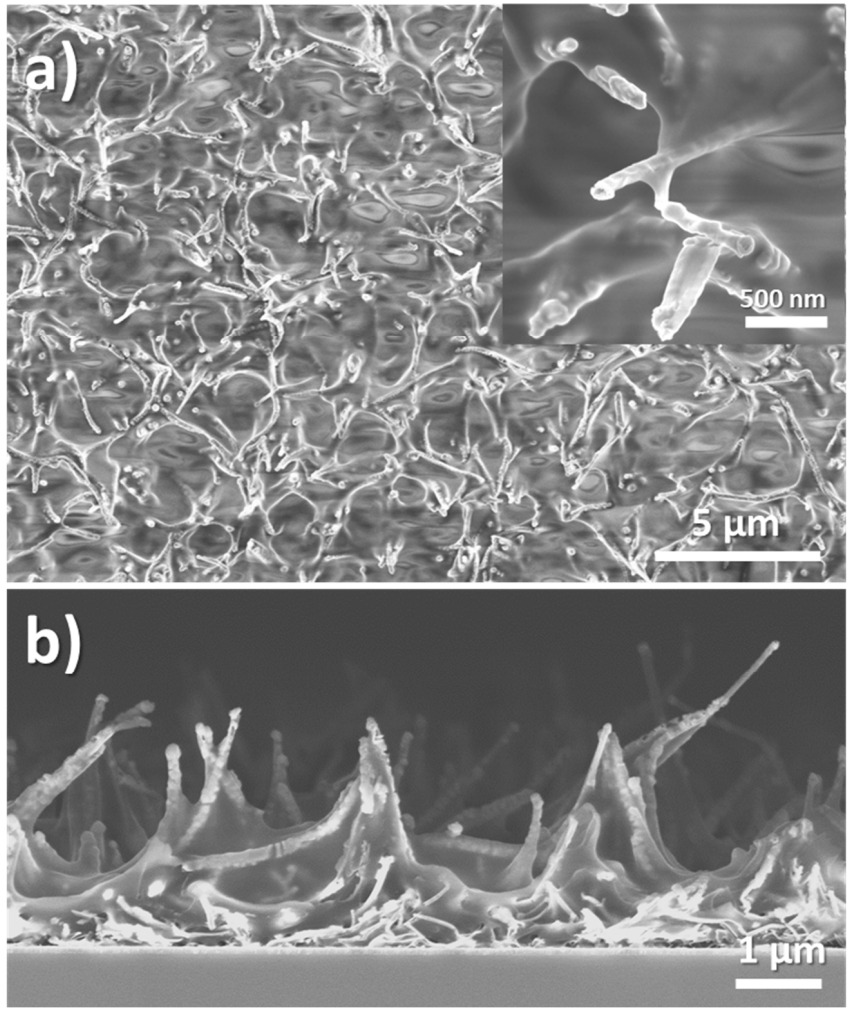

**Figure S12.** Top (a) and cross-sectional (b) SEM images of TiO2@MAPI NT encapsulated with PMMA layer. The inset in a) shows a high magnification of the unencapsulated tips of the NT.

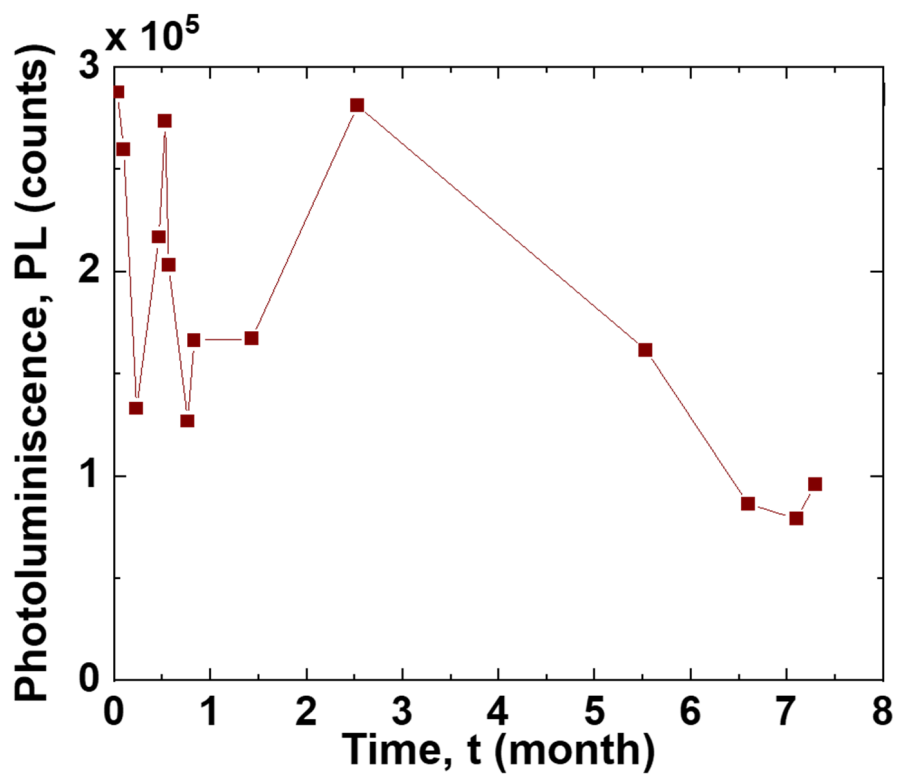

**Figure S13.** Photoluminescence of a TiO$_2$@MAPI NT sample encapsulated with PMMA under dark storage, measured during a period of 8 months.

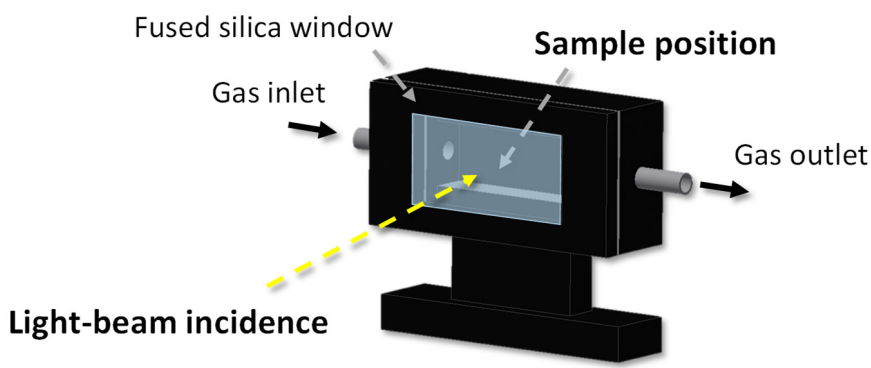

**Figure S14.** Schematic of the photoluminescence sample holder with atmosphere control used for stability tests.